# Generalized Theory of Förster-type Nonradiative Energy Transfer in Nanostructures with Mixed Dimensionality

Pedro Ludwig Hernández-Martínez[1,2], Alexander O. Govorov[3], and Hilmi Volkan Demir[1,2]

[1]LUMINOUS! Center of Excellence for Semiconductor Lighting and Display,
Physics and Applied Physics Division, School of Physical and Mathematical Sciences,
Microelectronics Division, School of Electrical and Electronics Engineering,
Nanyang Technological University, Singapore 639798, Singapore

[2]Department of Physics, Department of Electrical and Electronics Engineering,
UNAM - National Nanotechnology Research Center and Institute of Materials Science
and Nanotechnology, Bilkent University, Ankara 06800, Turkey

[3]Department of Physics and Astronomy, Ohio University, Athens, OH, 45701, USA

Abstract

Förster-type nonradiative energy transfer (NRET) is widely used, especially utilizing nanostructures in different combinations and configurations. However, the existing well-accepted Förster theory is only for the case of a single particle serving as a donor together with another particle serving as an acceptor. There are also other special cases previously studied; however, there is no complete picture and unified understanding. Therefore, there is a strong need for a complete theory that models Förster-type NRET for the cases of mixed dimensionality including all combinations and configurations. We report a generalized theory for the Förster-type NRET, which includes the derivation of the effective dielectric function due to the donor in different confinement geometries and the derivation of transfer rates distance dependencies due to the acceptor in different confinement geometries, resulting in a complete picture and understanding of the mixed dimensionality.

## I. INTRODUCTION

Semiconductor nanostructures have been studied extensively in the last decade [1,2,3]. Today nanotechnology offers means to assemble and study superstructures, e.g., those composed of nanocrystals and biomolecules [4,5,6]. For instance, by using a biomolecular linker, one can assemble crystalline nanoparticles and nanowires into complex structures with new physical and chemical properties [7,8,9,10]. Unique properties result from the quantum confinement and chemical composition of the building blocks of these superstructures (e.g., metal nanparticles, semiconductor nanocrystals, biomolecules). Furthermore, interactions between these unit elements lead to enhanced properties for the hybrid superstructure. One of the important mechanisms for strong interaction between the building blocks is the nonradiative energy transfer (NRET), also dubbed Förster-type resonance energy transfer (FRET) [11,12]. NRET results from the Coulomb interaction between excitons confined in the nanostructures. It is an efficient mechanism for coupling optically excited nanostructures. Such coupled superstructures can strongly change their physical properties. In particular, these changes can be observed in optical experiments of characterizing strongly packed structures. In the presence of optical excitation, the Coulomb (dipole-dipole) interaction results in the energy transfer between the elements (or building blocks) [7,13,14,15,16,17]. The resulting energy transfer effect is observed by the directional flow of excitons between the building blocks.

Numerous studies have previously been reported on NRET and related effects in systems of certain dimensionality. For example, several aspects of the energy transfer for nanoparticle-nanoparticle, nanoparticle-biomolecule, nanoparticle-nanorod, nanoparticle-surface, and quantum well-nanoparticle were discussed in Refs. 8,14, 18, 19, 20, 21, 22, 23, 24. Also, Refs. 25, 26, 27, 28, 29, 30, 31, 32, 33, 34 studied the energy transfer for the hybrid systems including nanoparticle-nanowire, nanoparticle-nanorod, nanoparticle-nanotube, nanowire-nanowire, and nanoparticle-nanosheet. Ref. 35 reported chemically-controlled NRET in nanoparticle composites. Nevertheless, to date a complete unified understanding on the modifications of NRET when using mixed dimensionality, with all possible combinations of quantum objects, e.g., nanoparticles (NPs); quantum wires, i.e., nanowires (NWs); and quantum wells (QWs) lacks. Although, the resulting NRET rates are fundamentally modified due to the mixed dimensionality, the distance dependency correlated to the dimension of the donor or the acceptor and their roles in modifying NRET have not been understood. However, understanding these modifications is essential to utilizing these nanostructures for high efficiency light generation and harvesting. Therefore, differentiating these effects with respect to the donor *vs.* the acceptor is critically important.

In this paper, a complete study of the generalized Förster-type NRET between nanostructures consisting of mixed dimensions in confinement (NP, NW, and QW) is presented. We investigate the modification of NRET mechanism with respect to the nanostructure serving as the donor *vs.* the acceptor, focusing on the rate's distance dependency and the role of the effective dielectric constant on NRET. In this work, the combinations of NW→NP, QW→NP, QW→NW, and NW→QW (where the donor→acceptor (D→A) denotes the energy transfer directed from the donor to the acceptor) were specifically considered because they have not previously been theoretically studied. Moreover, we obtain a set of analytical expressions for NRET in the cases mentioned above and derive generic expressions for the dimensionality involved to present a unified generalized picture of NRET. Finally, the asymptotic behavior of these equations and the comparisons between all possible cases are presented and discussed.

## II. DIPOLE POTENTIAL FOR EXCITON IN A NP, NW, AND QW

In this section, the analytical equations for the exciton electric potential inside and outside the nanostructures are obtained. For the long distance approximation, a set of convenient expressions for the outside electric potential are also derived. Moreover, we obtain the effective dielectric constant expressions in this limit.

### A. NP Case

The electric potential for an exciton in the $\alpha$-direction $(\alpha = x, y, z)$ inside a spherical NP, illustrated in Figure 1(a), is given by

$$\Phi_\alpha^{in} = \left(\frac{ed_{exc}}{\varepsilon_{NP}}\right)\frac{\hat{\mathbf{a}}\cdot\mathbf{r}}{r^3}\left(1 + \frac{2(\varepsilon_{NP}-\varepsilon_0)}{\varepsilon_{NP}+2\varepsilon_0}\frac{r^3}{R_{NP}^3}\right) \quad (1)$$

$$\Phi_\alpha^{out} = \left(\frac{ed_{exc}}{\varepsilon_{NP}}\right)\left(\frac{3\varepsilon_{NP}}{\varepsilon_{NP}+2\varepsilon_0}\right)\frac{\mathbf{r}\cdot\hat{\mathbf{a}}}{r^3} \quad (2)$$

where $ed_{exc}$ is the exciton dipole moment and $\varepsilon_{NP}$ and $\varepsilon_0$ are the NP and medium dielectric constants, respectively. The electric potential is the same in any direction because of the spherical symmetry of the NP. In the long distance approximation, the outside electric potential can be simplified as

$$\Phi_\alpha^{out} = \left(\frac{ed_{exc}}{\varepsilon_{eff}}\right)\frac{\mathbf{r}\cdot\hat{\mathbf{a}}}{r^3} \quad (3)$$

where $\varepsilon_{eff}$ is the effective dielectric constant given by

$$\varepsilon_{eff} = \frac{\varepsilon_{NP}+2\varepsilon_0}{3} \quad (4)$$

### B. NW Case

For the case of a long cylindrical NW, the electric potential for $\alpha$-exciton ($\alpha = x, y, z$), shown in Figure 1(a), can be written as the sum of the electric potential of the exciton inside the NW plus a second term to account for the boundary between the NW and the outside medium. Thus, the electric potential is expressed as

$$\Phi_\alpha^{in} = \Phi_\alpha + \sum_m \int \left(e^{im\varphi}e^{-iky}A_m^\alpha(k)I_m(|k|\rho)\right)dk \quad (5)$$

$$\Phi_\alpha^{out} = \Phi_\alpha + \sum_m \int \left(e^{im\varphi}e^{-iky}B_m^\alpha(k)K_m(|k|\rho)\right)dk \quad (6)$$

where $I_m(|k|\rho)$ and $K_m(|k|\rho)$ are the modified Bessel functions of order $m$; $\varphi$ is the angular component running from $0$ to $2\pi$, $k$ is the expansion along the cylinder axis $[k\in(-\infty,\infty)]$; and $\Phi_\alpha$ is the electric potential for an $\alpha$-exciton inside the NW. After applying the boundary conditions at the surface of the NW, $A_m^\alpha$ and $B_m^\alpha$ coefficients are found to be

$$A_m^\alpha(k) = \left(\frac{K_m(|k|R_{NW})}{I_m(|k|R_{NW})}\right)B_m^\alpha(k) \quad (7)$$

$$B_m^\alpha(k) = \frac{\frac{2}{|k|}(\varepsilon_0 - \varepsilon_{NW})g_m^\alpha(|k|)}{\varepsilon_{NW}\left(\frac{K_m(|k|R_{NW})}{I_m(|k|R_{NW})}\right)\mathcal{I}_m(|k|R_{NW}) + \varepsilon_0 \mathcal{K}_m(|k|R_{NW})} \tag{8}$$

where $\mathcal{I}_m(|k|R_{NW}) = I_{m-1}(|k|R_{NW}) + I_{m+1}(|k|R_{NW})$, $\mathcal{K}_m(|k|R_{NW}) = K_{m-1}(|k|R_{NW}) + K_{m+1}(|k|R_{NW})$, and $g_m^\alpha(|k|)$ is defined as

$$g_m^\alpha(|k|) = \frac{1}{(2\pi)^2}\int_0^{2\pi}\int_{-\infty}^{\infty} d\varphi dy\, e^{-im\varphi} e^{iky}\left[\frac{\partial \Phi_\alpha}{\partial \rho}\right]_{\rho=R_{NW}} \tag{9}$$

The outside electric potential, in the long distance approximation, for an exciton in the α-direction, is simplified as

$$\Phi_\alpha^{out} = \left(\frac{ed_{exc}}{\varepsilon_{eff}}\right)\frac{\mathbf{r}\cdot\hat{\mathbf{\alpha}}}{r^3} \tag{10}$$

where $\varepsilon_{eff}$ is the effective dielectric constant. $\varepsilon_{eff}$ is given by Eq. (11) if the exciton is along the cylinder main axis and by Eq. (12) when the exciton is perpendicular to the cylinder main axis.

$$\varepsilon_{eff} = \varepsilon_0 \tag{11}$$

$$\varepsilon_{eff} = \frac{\varepsilon_{NW} + \varepsilon_0}{2} \tag{12}$$

### C. QW Case

A thin QW embedded in a semi-infinite dielectric medium is considered as depicted in Figure 1(a). In a similar manner as the previous case, the electric potential can be written as the sum of the electric potential of the exciton inside the dielectric medium plus a second term to include the change between the dielectric and outside media. Therefore, the electric potential in the cylindrical coordinates for a α-exciton ($\alpha = x, y, z$) is written as

$$\Phi_\alpha^{in} = \Phi_\alpha + \sum_m \int_0^\infty kdk\, e^{-im\varphi} J_m(k\rho) A_m^\alpha(k) \cosh(kz) \tag{13}$$

$$\Phi_\alpha^{out} = \Phi_\alpha + \sum_m \int_0^\infty kdk\, e^{-im\varphi} J_m(k\rho) B_m^\alpha(k) Exp(-k|z|) \tag{14}$$

where $J_m(k\rho)$ is the Bessel function of order m, $\varphi$ is the angular component running from 0 to $2\pi$, k is the expansion along the cylinder radius, and the electric potential for an α-exciton inside the QW is $\Phi_\alpha$. The boundary conditions at the surface of the QW yield the coefficients $A_m^\alpha$ and $B_m^\alpha$

$$A_m^\alpha(k) = \left(\frac{\exp(-|k|L_{QW})}{\cosh(|k|L_{QW})}\right) B_m^\alpha(k) \tag{15}$$

$$B_m^\alpha(k) = \frac{(\varepsilon_0 - \varepsilon_{QW}) h_m^\alpha(|k|)}{k(\varepsilon_{QW}\tanh(|k|L_{QW}) + \varepsilon_0) e^{-|k|L_{QW}}} \tag{16}$$

where $h_m^\alpha(|k|)$ is defined as

$$h_m^\alpha(|k|) = \frac{1}{(2\pi)} \int_0^{2\pi}\int_0^\infty d\varphi\rho d\rho e^{im\varphi} J_m(k\rho)\left[\frac{\partial \Phi_\alpha}{\partial z}\right]_{z=L_{QW}} \tag{17}$$

The outside electric potential for an exciton in the *α*-direction, in the long distance approximation, is

$$\Phi_z^{out} = \left(\frac{ed_{exc}}{\varepsilon_{eff}}\right)\frac{\mathbf{r}\cdot\hat{\boldsymbol{\alpha}}}{r^3} \tag{18}$$

where $\varepsilon_{eff}$ is the effective dielectric constant giving by for all *α*-direction

$$\varepsilon_{eff} = \varepsilon_0 \tag{19}$$

A summary for the effective dielectric constant in the long distance approximation is given in Table 1, which shows the symmetry and screening factors for the electric potential in each case. This screening factor is the result of the interface between the nanostructure (NP, NW, and QW) and the outside medium. As expected, the screening factor for the NP case is the same for an exciton in the *x*-, *y*- and *z*-directions because of the spherical symmetry. In the cylindrical symmetry (NW case), an exciton in the cylindrical main axis does not have any screening factor because the exciton dipole moment is perpendicular to the cylinder's normal surface vector. However, an exciton perpendicular to the cylindrical main axis has a screening factor as given in Table 1. The reason for this screening factor stems from that fact that the exciton dipole moment is parallel to the cylinder's normal surface vector. In the QW case, we do not observe any screening factor for an exciton in the *x*-, *y*- or *z*-direction. This can be explained due to the fact that we choose the cylinder axis along the *z* axis and make the limit $\rho \to \infty$ to account for an infinite plane (QW). It is worth mentioning that we consider an infinite cylinder for the NW case and an infinite plane for the QW embedded in a semi-infinite dielectric medium. Note that Table 1 follows the geometries sketched in Figure 1.

**Table 1:** Effective dielectric constant expressions for the cases of NP, NW and QW in the long distance approximation. This table follows the geometries given in Figure 1.

| α-direction | NP | NW | QW |
|---|---|---|---|
| x | $\varepsilon_{eff_D} = \dfrac{\varepsilon_{NP_D} + 2\varepsilon_0}{3}$ | $\varepsilon_{eff_D} = \dfrac{\varepsilon_{NW} + \varepsilon_0}{2}$ | $\varepsilon_{eff} = \varepsilon_0$ |
| y | $\varepsilon_{eff_D} = \dfrac{\varepsilon_{NP_D} + 2\varepsilon_0}{3}$ | $\varepsilon_{eff} = \varepsilon_0$ | $\varepsilon_{eff} = \varepsilon_0$ |
| z | $\varepsilon_{eff_D} = \dfrac{\varepsilon_{NP_D} + 2\varepsilon_0}{3}$ | $\varepsilon_{eff_D} = \dfrac{\varepsilon_{NW} + \varepsilon_0}{2}$ | $\varepsilon_{eff} = \varepsilon_0$ |

Figure 1 depicts the total and long distance approximation electric potentials for a *z*-exciton along the *z* axis. Figure 1(b) shows both the total and long distance approximation electric potentials for a *z*-exciton inside an NP. It is observed that both electric potentials overlap each other because of the spherical symmetry of the nanostructure (Eq. (1), Eq. (2), and Eq. (3)). In a similar manner, the total and long distance approximation electric potentials for a *z*-exciton in a long NW are depicted in Figure 1(c). In close proximity to the NW surface, the long distance approximation underestimates the exciton electric potential, as it is shown in Figure 1(c). In the QW case, the long distance approximation overestimates the exciton electric potential in the close proximity to the QW surface (Figure 1(d)). This is an opposite effect compared to the NW case. In all cases, at long distances the total electric potential converges to the long distance approximation (Figure 1(b), 1(c) and 1(d)).

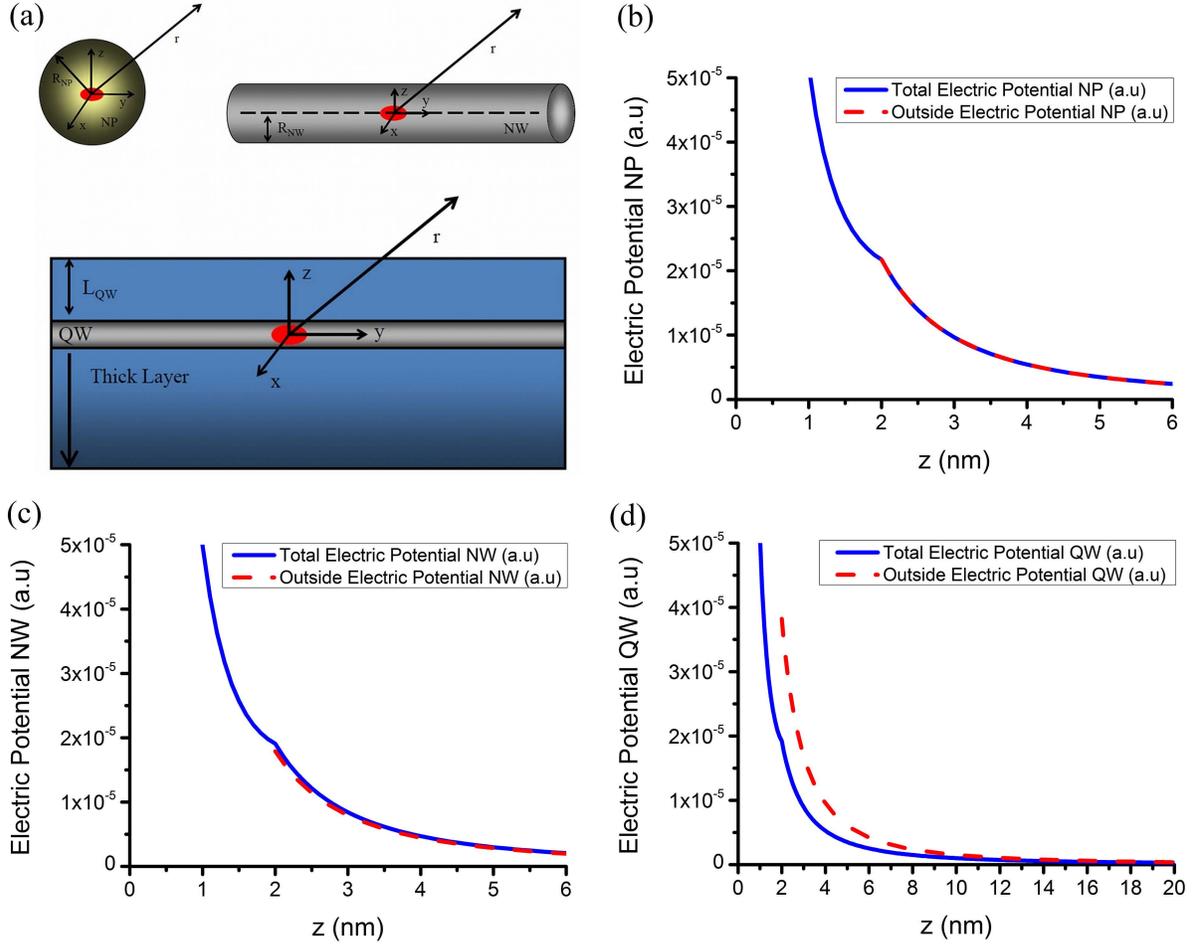

**Figure 1:** a) Schematic for an exciton in a NP, NW, and QW. Red circle represents an exciton in the α-direction. $R_{NP(NW)}$ is the NP (NW) radius. $L_{QW}$ is the QW capping layer thickness. Total and long distance approximation electric potentials for the *z*-exciton inside b) an NP; c) an NW; and d) a QW.

## III. THEORETICAL FORMALISM FOR FÖRSTER-TYPE NONRADIATIVE ENERGY TRANSFER

In this section, we outline the macroscopic approach to the problem of dipole-dipole energy transfer. We restrict ourselves to the case of a single electron-hole pair (exciton) in the donor nanostructure. Moreover, we consider only two states $|0\rangle$ - the ground state and $|exc\rangle$ - the excited state. These states are considered using simplified wavefunctions such that we consider excitonic states without the mixing between the heavy- and light-hole states. Furthermore, the spin part is not considered in our model.

NRET is a directional process, which is initiated by an absorbed photon in a donor generating an exciton in a higher excited state and then the exciton relaxing very fast to the first excited state by higher order processes. This exciton can subsequently be either recombined (through a radiative or nonradiative means) or transferred to an acceptor because of the Coulomb interaction

7 | P a g e

between dipoles in the D-A pair. If the exciton is transferred, it will occupy a higher excited state in the acceptor and relax (very fast) to its first excited state to finally recombine through a radiative or nonradiative process. NRET occurs only when the donor possesses a greater or equal band gap compared to the acceptor. The diagram for the energy transfer process is shown in Figure 2.

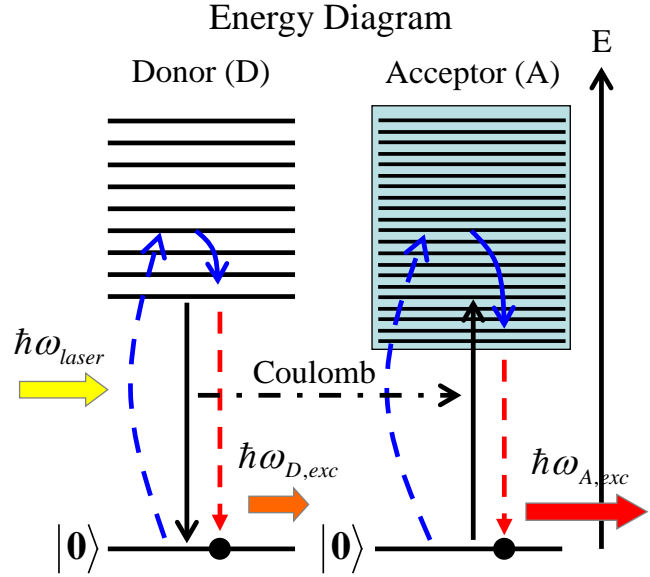

**Figure 2:** Energy diagram for the directional process of exciton transfer from the donor to the acceptor. Blue dash lines represent the absorption process of the nanostructure (donor/acceptor). Blue solid lines denote fast relaxation process. Red dash lines show light emission process (relaxation from the lowest excited state to ground state). Black solid lines represent the energy transfer from the donor to the acceptor. Horizontal solid black line illustrates the Coulomb interaction between the donor and the acceptor[27].

The probability of an exciton transfer from a donor nanostructure (donor) to an acceptor nanostructure (acceptor) is given by the *Fermi's Golden rule* (Eq. (20)).

$$\gamma_{trans} = \frac{2}{\hbar}\left\{\sum_{f}\left|\langle f_{exc};0_{exc}|\hat{V}_{int}|i_{exc};0_{exc}\rangle\right|^2 \delta(\hbar\omega_{exc}-\hbar\omega_f)\right\} \quad (20)$$

where $|i_{exc};0_{exc}\rangle$ is the initial state with an exciton in the donor and zero exciton in the acceptor; $|f_{exc};0_{exc}\rangle$ is the final state with an exciton in the acceptor and zero exciton in the donor; $\hat{V}_{int}$ is the exciton Coulomb interaction operator; and $\hbar\omega_{exc}$ is the exciton's energy. Neglecting the coherent coupling between excitons, the initial and final states can be written as $|i_{exc};0_{exc}\rangle = |i_{exc}\rangle|0_{exc}\rangle$ and $|f_{exc};0_{exc}\rangle = |f_{exc}\rangle|0_{exc}\rangle$, respectively, and the Fermi's Golden rule can be approximated by

$$\gamma_{trans} \approx \frac{2}{\hbar}\left\{\sum_{f}\left|\langle f_{exc}|\hat{U}_{int}|0_{exc}\rangle\right|^2 \delta(\hbar\omega_{exc}-\hbar\omega_f)\right\} \quad (21)$$

where $\hat{U}_{int} = \langle 0_{exc} | \hat{V}_{int} | i_{exc} \rangle$ is the potential energy created by the exciton. With this approximation, the Fermi's Golden rule can be simplified with the used of the *fluctuation dissipation theorem* (FDT) [36] together with the QD formalism developed in ref. [37] and [27]. The final expression for the transfer rate is given by

$$\gamma_{trans} = \frac{2}{\hbar} \text{Im} \left[ \int dV \left( \frac{\varepsilon_A(\omega)}{4\pi} \right) \mathbf{E}_{in}(\mathbf{r}) \cdot \mathbf{E}_{in}^*(\mathbf{r}) \right] \quad (22)$$

where the integration is taken over the acceptor volume, $\varepsilon_A(\omega)$ is the dielectric function of the acceptor, and $\mathbf{E}_{in}(\mathbf{r})$ includes the effective electric field created by an exciton at the donor side. The electric field is calculated with $\mathbf{E}(\mathbf{r}) = -\nabla \Phi(\mathbf{r})$ and the electric potential $\Phi(\mathbf{r})$ is given by

$$\Phi_\alpha(\mathbf{r}) = \left( \frac{ed_{exc}}{\varepsilon_{effD}} \right) \frac{(\mathbf{r} - \mathbf{r}_0) \cdot \hat{\boldsymbol{\alpha}}}{|\mathbf{r} - \mathbf{r}_0|^3} \quad (23)$$

where $ed_{exc}$ is the dipole moment of the exciton and $\varepsilon_{effD}$ is the effective dielectric constant of the donor, which depends on the geometry and the exciton dipole direction, $\alpha = x, y, z$. Furthermore, the average NRET rate (at room temperature) is calculated as

$$\gamma_{trans} = \frac{\gamma_{x,trans} + \gamma_{y,trans} + \gamma_{z,trans}}{3} \quad (24)$$

where $\gamma_{\alpha,trans}$ is the transfer rate for the α-exciton ($\alpha = x, y, z$).

## A. NW→NP AND QW→NP ENERGY TRANSFER RATES

The NRET rate analytical equations, when the donor is a NW or a QW and the acceptor is a NP (Figure 3(a) and 3(b), respectively), are derived. Moreover, in the long distance approximation, we obtain simplified expressions for the transfer rate for these cases (NW→NP and QW→NP). Assuming that the donor size is smaller than the separation distance between the D-A pair and using the spherical symmetry of the acceptor, the total electric potential for the acceptor can be written as

$$\Phi_\alpha^{out}(r,\theta,\phi) = \Phi_\alpha(r,\theta,\phi) + \sum_{l,m} \frac{B_{l,m}^\alpha}{r^{l+1}} Y_{l,m}(\theta,\phi) \quad (25)$$

$$\Phi_\alpha^{in}(r,\theta,\phi) = \sum_{l,m} A_{l,m}^\alpha r^l Y_{l,m}(\theta,\phi) \quad (26)$$

where $\Phi_\alpha(r,\theta,\phi)$ is the electric potential of the exciton in the donor; $Y_{l,m}(\theta,\phi)$ are the spherical harmonics; and $A_{l,m}^{\alpha}$ and $B_{l,m}^{\alpha}$ are the coefficients determined by the boundary conditions at the surface of the NP ($\Phi_{\alpha}^{in}(r=R_{NP_A},\theta,\phi)=\Phi_{\alpha}^{out}(r=R_{NP_A},\theta,\phi)$ and $\varepsilon_{in}\left(\frac{\partial \Phi_{\alpha}^{in}(r,\theta,\phi)}{\partial r}\right)_{r=R_{NP_A}} = \varepsilon_{out}\left(\frac{\partial \Phi_{\alpha}^{out}(r,\theta,\phi)}{\partial r}\right)_{r=R_{NP_A}}$). $\varepsilon_{in(out)}$ is the dielectric function inside (outside) the acceptor. After applying the boundary conditions we obtain

$$A_{l,m}^{\alpha} = \frac{B_{l,m}^{\alpha}}{R_{NP_A}^{2l+1}} + \frac{f_{l,m}^{\alpha}}{R_{NP_A}^{l}} \qquad (27)$$

$$B_{l,m}^{\alpha} = \frac{R_{NP_A}^{l+2}\left(\varepsilon_{out}g_{l,m}^{\alpha} - l\varepsilon_{in}\frac{f_{l,m}^{\alpha}}{R_{NP_A}}\right)}{l\varepsilon_{in} + (l+1)\varepsilon_{out}} \qquad (28)$$

with $f_{l,m}^{\alpha}$, $g_{l,m}^{\alpha}$, which are given by

$$f_{l,m}^{\alpha} = \int_0^{2\pi}\int_0^{\pi}\left[\Phi_{\alpha}(r,\theta,\phi)\right]_{r=R_{NP_A}} Y_{l,m}^{*}(\theta,\phi)\sin(\theta)d\theta d\phi \qquad (29)$$

$$g_{l,m}^{\alpha} = \int_0^{2\pi}\int_0^{\pi}\left[\frac{\partial \Phi_{\alpha}(r,\theta,\phi)}{\partial r}\right]_{r=R_{NP_A}} Y_{l,m}^{*}(\theta,\phi)\sin(\theta)d\theta d\phi \qquad (30)$$

and $\varepsilon_{out} = \varepsilon_0$ is the dielectric constant of the medium, and $\varepsilon_{in} = \varepsilon_{NP_A}$ is the dielectric function of the acceptor. Thus, the energy transfer rate is

$$\gamma_{\alpha,trans} = \frac{2}{\hbar}\text{Im}\left[\varepsilon_{NP_A}(\omega)\left(\frac{1}{4\pi}\right)\sum_{l,m}\left|A_{l,m}^{\alpha}\right|^2 \cdot l \cdot R_{NP_A}^{2l+1}\right] \qquad (31)$$

where $A_{l,m}^{\alpha}$ is given by Eq. (27). This is a general expression valid under the assumption mentioned above. From Eq. (31), it is observed that the distance dependency for the transfer rate is given by the coefficient $A_{l,m}^{\alpha}$. The asymptotic behavior (in the long distance limit) for the transfer rate in the dipole approximation for NW→NP and QW→NP is derived. In these cases, we assume that the donor size is small compared to the separation distance $d$. Under this condition, the transfer rate ($\gamma_{\alpha,trans}$) is

$$\gamma_{\alpha,trans} = \frac{2}{\hbar}b_{\alpha}\left(\frac{ed_{exc}}{\varepsilon_{eff_D}}\right)^2 \frac{R_{NP_A}^3}{d^6}\cos^6(\theta_0)\left|\frac{3\varepsilon_0}{\varepsilon_{NP_A}(\omega_{exc})+2\varepsilon_0}\right|^2 \text{Im}\left[\varepsilon_{NP_A}(\omega_{exc})\right] \qquad (32)$$

where $b_{\alpha} = \frac{1}{3}, \frac{1}{3}, \frac{4}{3}$ for $\alpha = x,y,z$, respectively; $\theta_0$ is the angle between $d$ and $\mathbf{r}$; $\varepsilon_{eff_D}$ is the effective dielectric constant for the exciton in the donor. For the NW-to-NP case, the effective dielectric constant is $\varepsilon_{eff_D} = \varepsilon_0$ (Eq. (11)) for $\alpha = y$ (parallel to the cylindrical axis) and $\varepsilon_{eff_D} = \frac{\varepsilon_{NW} + \varepsilon_0}{2}$ (Eq. (12)), $\alpha = x, z$ (perpendicular to the cylindrical axis). For the QW-to-NP case, it is equal to $\varepsilon_{eff_D} = \varepsilon_0$ (Eq. (19)) for $\alpha = x, y, z$. The NRET rates are proportional to the imaginary part of the acceptor dielectric constant. Thus, an acceptor with high absorption (large $\text{Im}|\varepsilon_{NP_A}(\omega)|$) will have higher transfer rates. Moreover, the transfer rate strongly depends on the distance and $\theta_0$. In particular for the angle dependency, the main contribution comes from small $\theta_0$ and decreases very fast as $\theta_0$ increases. It is important to notice that the transfer rate in these cases (NW-to-NP and QW-to-NP) follows the same distance dependency as the NP-to-NP transfer rate, which is $\gamma \propto d^{-6}$ [38]. These results suggest that the NRET rates are dictated by the acceptor's dimensionality, but not the donor's.

To illustrate the NRET rate, we present the average NRET rate in the long distance approximation as a function of distance for CdTe D-A pair in Figure 3(c) and 3(d). Here, we consider the donor to be an NW or a QW and the acceptor to be an NP. In this plot, we assume that the acceptor exciton emission is at $\lambda = 582$ nm. In addition, the acceptor dielectric function is taken from Ref. 39. Figure 3(c) shows the energy transfer rate for the QW-to-NP case. Figure 3(d) depicts the contour profile plot for the QW-to-NP transfer rate. The top panel on Figure 3(d) illustrates the energy transfer rate as a function of the distance at a fixed angle. Blue curve represents the case at $\theta_0 = 0$, and wine curve, at $\theta_0 = \pi / 6$. The right panel on Figure 3(d) shows the transfer rate as a function of angle at a fixed distance. Red curve represents the case at $d = 3.3$ nm, and the green curve, at $d = 4.0$ nm. From Figure 3(c) and 3(d), the strong distance dependency of the transfer rate (Eq. (32)) is observed. Therefore, the main contribution for the energy transfer from a QW(NW) to an NP comes at short distances and small angles.

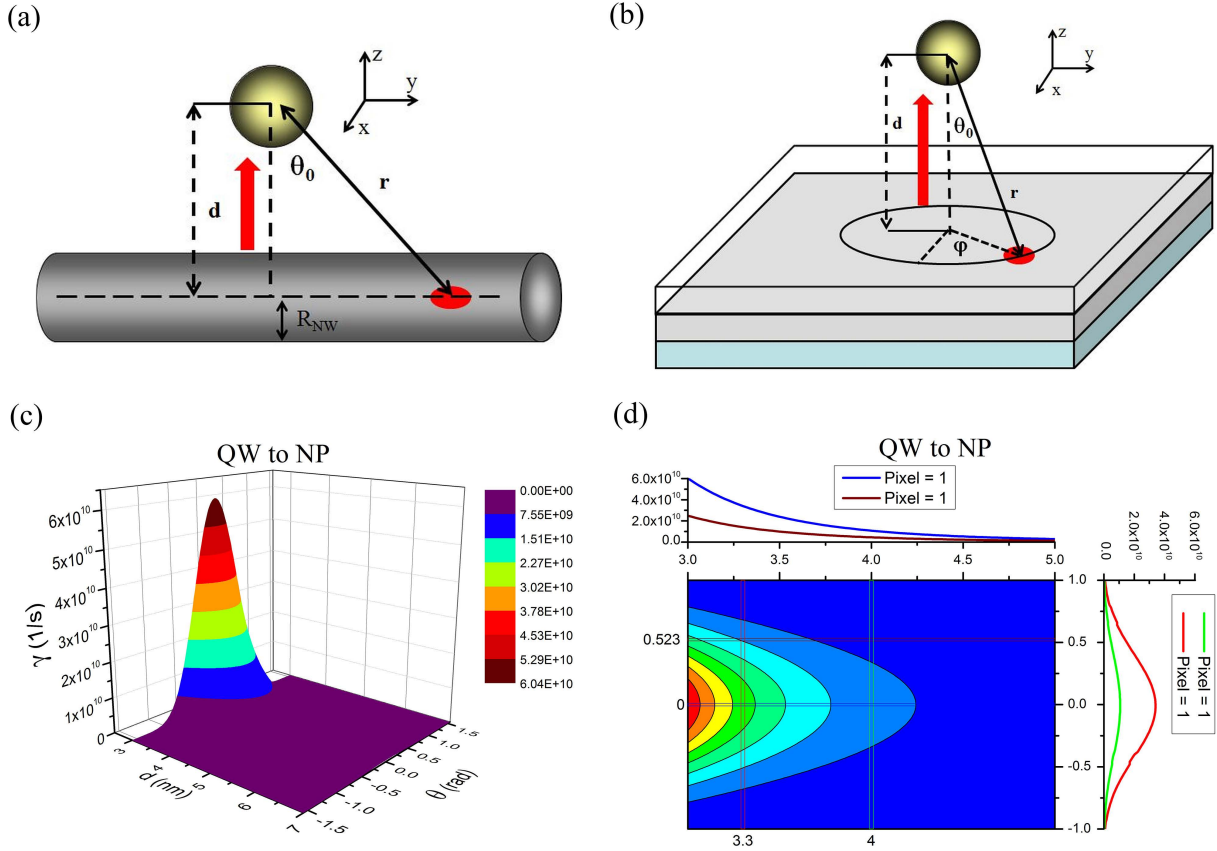

**Figure 3:** Schematic for the energy transfer of a) NW→NP and b) QW→NP. Red arrows show the energy transfer direction. Red circles represent an exciton in the α-direction. *d* is the separation distance. $\theta_0$ is the azimuthal angle between ***d*** and ***r***. $\varphi$ is the radial angle. c) Average NRET rate for the CdTe D-A QW→NP pair as a function of the distance and angle. d) Contour profile map for the average NRET rate for the CdTe D-A QW→NP pair, with the top panel at a fixed angle and right panel at a fixed distance.

## B. QW→NW ENERGY TRANSFER RATE

We derive an analytical equation for the NRET rate when the donor is a QW and the acceptor is an NW (Figure 4(a)). The simplified expression for NRET rate in the long distance approximation is also obtained. Similar to the previous case, we assume that the donor size is small compared to the D-A separation distance *d*. Taking advantage of the cylindrical symmetry of the acceptor, the total electric potential for the acceptor can be written as

$$\Phi_\alpha^{out}(\rho,\phi,z) = \Phi_\alpha(\rho,\phi,z) + \sum_m \int_{-\infty}^{\infty} dk\, e^{-ikz} B_m^\alpha(k) K_m(|k|\rho) e^{im\phi} \tag{33}$$

$$\Phi_\alpha^{in}(\rho,\phi,z) = \sum_m \int_{-\infty}^{\infty} dk\, e^{-ikz} A_m^\alpha(k) I_m(|k|\rho) e^{im\phi} \tag{34}$$

where $\Phi_\alpha(\rho,\phi,z)$ is the electric potential of the exciton in the donor; $I_m(|k|\rho)$ and $K_m(|k|\rho)$ are the modified Bessel functions; and $A_m^\alpha(k)$, $B_m^\alpha(k)$ are the coefficients given by Eq. (35) and Eq. (36). These coefficients are determined by the boundary conditions $\Phi_\alpha^{in}(\rho = R_{NW_A},\phi,z) = \Phi_\alpha^{out}(\rho = R_{NW_A},\phi,z)$ and $\varepsilon_{NW}\left(\frac{\partial \Phi_\alpha^{in}(\rho,\phi,z)}{\partial \rho}\right)_{\rho=R_{NW_A}} = \varepsilon_0\left(\frac{\partial \Phi_\alpha^{out}(\rho,\phi,z)}{\partial \rho}\right)_{\rho=R_{NW_A}}$, where $\varepsilon_{NW(0)}$ is the NW and outside medium dielectric function, respectively.

$$A_m^\alpha(k) = \frac{K_m(|k|R_{NW_A})}{I_m(|k|R_{NW_A})} B_m^\alpha(k) + \frac{f_m^\alpha(|k|)}{I_m(|k|R_{NW_A})} \tag{35}$$

$$B_m^\alpha(k) = \frac{\frac{2}{|k|}\varepsilon_0 g_m^\alpha(|k|) - \varepsilon_{NW} \frac{\mathcal{I}_m(|k|R_{NW_A})}{I_m(|k|R_{NW_A})} f_m^\alpha(|k|)}{\varepsilon_{NW}\left(\frac{\mathcal{I}_m(|k|R_{NW_A})}{I_m(|k|R_{NW_A})}\right) K_m(|k|R_{NW_A}) + \varepsilon_0 \mathcal{K}_m(|k|R_{NW_A})} \tag{36}$$

with $\mathcal{I}_m(|k|R_{NWA}) = I_{m+1}(|k|R_{NWA}) + I_{m-1}(|k|R_{NWA})$, $\mathcal{K}_m(|k|R_{NWA}) = K_{m+1}(|k|R_{NWA}) + K_{m-1}(|k|R_{NWA})$, and $f_m^\alpha$, and $g_m^\alpha$ are given by

$$f_m^\alpha = \frac{1}{(2\pi)^2} \int_0^{2\pi}\int_{-\infty}^{\infty} \left[\Phi_\alpha(\rho,\phi,z)\right]_{\rho=R_{NW_A}} e^{ikz} e^{-im\phi}\, dz d\phi \tag{37}$$

$$g_m^\alpha = \frac{1}{(2\pi)^2} \int_0^{2\pi}\int_{-\infty}^{\infty} \left[\frac{\partial \Phi_\alpha(\rho,\phi,z)}{\partial \rho}\right]_{\rho=R_{NW_A}} e^{ikz} e^{-im\phi}\, dz d\phi \tag{38}$$

The energy transfer rate is written as

$$\gamma_{\alpha,trans} = \frac{2}{\hbar} \text{Im}\left[\frac{\varepsilon_{NW_A}(\omega_{exc})}{4\pi}\right] (2\pi)^2 \sum_m \int_{-\infty}^{\infty} dk\, |A_m^\alpha(|k|)|^2 \times$$

$$\left(\frac{|k|^2}{4} \int_0^{R_{NW_A}} |\mathcal{I}_m(|k|\rho)|^2 \rho d\rho + m^2 \int_0^{R_{NW_A}} |I_m(|k|\rho)|^2 \frac{1}{\rho} d\rho + |k|^2 \int_0^{R_{NW_A}} |I_m(|k|\rho)|^2 \rho d\rho\right) \tag{39}$$

where $A_m^\alpha(k)$ (Eq. (35)) gives the distance dependency to the NRET rate. Eq. (39) is a general expression, again which is valid under the assumptions mentioned above. In the long distance approximation, the transfer rate equation for the QW-to-NW case is

$$\gamma_{\alpha,trans} = \frac{2}{\hbar}\left(\frac{ed_{exc}}{\varepsilon_{eff_D}}\right)^2 \left(\frac{3\pi}{32}\right)\frac{R_{NW_A}^2}{d^5}\cos^5(\theta_0)\left(a_\alpha + b_\alpha\left|\frac{2\varepsilon_0}{\varepsilon_{NW_A}(\omega_{exc})+\varepsilon_0}\right|^2\right)\text{Im}\left[\varepsilon_{NW_A}(\omega_{exc})\right] \quad (40)$$

where $a_\alpha = 0, \frac{9}{16}, \frac{15}{16}$; $b_\alpha = 1, \frac{15}{16}, \frac{41}{16}$ for $\alpha = x, y, z$, respectively; $d$ is the center-to-center distance between the donor and the acceptor; $\theta_0$ is the angle between $d$ and $\mathbf{r}$; and $\varepsilon_{eff_D}$ is the effective dielectric constant for the exciton in the donor, which is equal to $\varepsilon_{eff_D} = \varepsilon_0$ (Eq. (19)) for $\alpha = x, y, z$. As expected, the asymptotic behavior for the NRET rate of the QW→NW case follows $\gamma \propto d^{-5}$. This result is similar to the NP-to-NW and NW-to-NW cases, as reported in previous literature [27] and [28], respectively. Similar to the previous section, the NRET rates strongly depend on the distance and $\theta_0$, and a similar analysis can be made. In addition, akin to the section above, Figure 4(b) and 4(c) depict the average NRET rate for a CdTe D-A pair as a function of the distance and $\theta_0$, when the donor is a QW and the acceptor is an NW. For this plot, we assume that the acceptor exciton emission is at $\lambda = 610$ nm and the acceptor dielectric function is taken from Ref. 39. Figure 4(c) shows the contour profile map for the QW-to-NW transfer rate. The top panel in Figure 4(c) illustrates the energy transfer rate as a function of the distance at a fixed angle. Blue curve represents the situation at $\theta_0 = 0$, and wine curve, at $\theta_0 = \pi/6$. The right panel in Figure 3(c) shows the transfer rate as a function of the angle at a fixed distance. Red curve represents the behavior at $d = 3.3$ nm, and the green curve, at $d = 4.0$ nm. Figure 4(d) gives the semi-log plot for the rates as a function of the distance at $\theta_0 = 0$. From Figure 4(b), 4(c), and 4(d), the strong distance dependency of the transfer rate (Eq. (40)), can be seen. Therefore, similar to the previous section, the main contribution for the energy transfer from a QW to an NW comes at short distances and small angles.

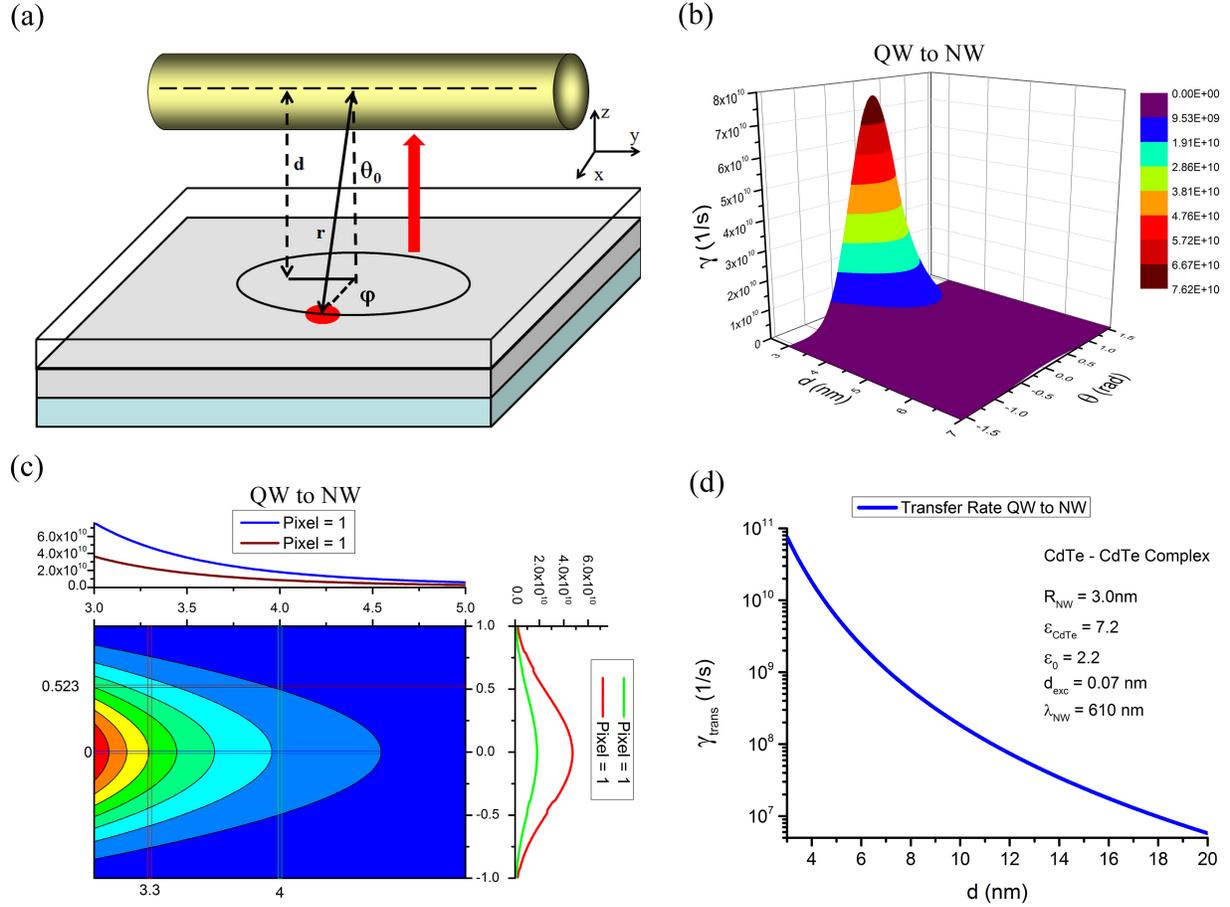

**Figure 4:** a) Schematic for the energy transfer of QW→NW. Red arrows show the energy transfer direction. Red circles represent an exciton in the α-direction. $d$ is the separation distance. $\theta_0$ is the azimuthal angle between $\boldsymbol{d}$ and $\boldsymbol{r}$. $\varphi$ is the radial angle. b) Average NRET rate for the CdTe D-A QW→NW pair as a function of the distance and angle. c) Contour profile map for the average NRET rate of QW→NW, with the top panel at a fixed angle, and right panel at a fixed distance. d) Average NRET rate for CdTe D-A pair with $\theta_0 = 0$. This plot illustrates the NRET rate distance dependency for the QW→NW case.

## C. NW→QW ENERGY TRANSFER

The analytical equation for the NRET rate when the donor is an NW and the acceptor is a QW (Figure 5(a)) is derived. Moreover, the simplified expression for the NRET rate in the long distance approximation is also obtained. Similar to the previous cases, we assume that the donor size is small compared to the D-A separation distance $d$. Furthermore, we consider a symmetric structure, consisting of a semiconductor QW of thickness $L_w$ between two barriers of dielectric function $\varepsilon_{QW}$. One barrier has a film thickness $L_l$, while the other barrier is considered to be very thick (we assume that this barrier is semi-infinite). The donor nanostructure is placed in front of the barrier with thickness $L_l$ and we solve the problem for the case where the QW is very thin ($L_w \ll L_l$). Under these assumptions, the electric potential inside the barrier is

$$\Phi_{in}(\mathbf{r}) = \left(\frac{2\varepsilon_0}{\varepsilon_{QW} + \varepsilon_0}\right)\Phi_\alpha(\mathbf{r}) \tag{41}$$

where $\varepsilon_0$ is the dielectric constant outside the barrier and $\Phi_\alpha$ is the electric potential of an $\alpha$-exciton in the QW. Therefore, the transfer rate is

$$\gamma_{\alpha,trans} = \frac{2}{\hbar}\left|\frac{2\varepsilon_0}{\varepsilon_{QW_A} + \varepsilon_0}\right|^2 \text{Im}\left[\int dV \left(\frac{\varepsilon_{QW_A}(\omega)}{4\pi}\right) \mathbf{E}_\alpha(\mathbf{r}) \cdot \mathbf{E}_\alpha^*(\mathbf{r})\right] \tag{42}$$

where $E_\alpha(\mathbf{r})$ is the electric field created by an $\alpha$-exciton in the donor. By using the assumption that the QW is very thin ($L_w \ll L_l$), the energy transfer rate becomes

$$\gamma_{\alpha,trans} = \frac{2}{\hbar}\left|\frac{2\varepsilon_0}{\varepsilon_{QW_A} + \varepsilon_0}\right|^2 \text{Im}\left[\int_{QW_A} dS \left(\frac{\varepsilon_{QW_A}(\omega)}{4\pi}\right) \mathbf{E}_\alpha(\mathbf{r}) \cdot \mathbf{E}_\alpha^*(\mathbf{r})\right] \tag{43}$$

where the integration is taken over the surface of the QW. In particular, we obtain the analytical expression in the long distance approximation for NW→QW. In this case, we assume $d_b \gg L_W$ where $d_b$ is the distance from the center of the donor to the dielectric barrier. Under these conditions, $\gamma_{\alpha,trans}$ becomes

$$\gamma_{\alpha,trans} = \frac{2}{\hbar}b_\alpha \left(\frac{ed_{exc}}{\varepsilon_{eff_D}}\right)^2 \frac{1}{d^4}\left|\frac{2\varepsilon_0}{\varepsilon_{QW_A} + \varepsilon_0}\right|^2 \text{Im}[\varepsilon_{QW_A}(\omega_{exc})] \tag{44}$$

where $b_\alpha = \frac{3}{16}, \frac{3}{16}, \frac{3}{8}$ for $\alpha = x, y, z$, respectively; $d = d_b + L_l$ is the distance between the donor and the acceptor; and $\varepsilon_{eff_D}$ is the effective dielectric constant given by $\varepsilon_{eff_D} = \varepsilon_0$ (Eq. (11)) for $\alpha = y$ (parallel to the cylindrical axis) and $\varepsilon_{eff_D} = \frac{\varepsilon_{NW} + \varepsilon_0}{2}$ (Eq. (12)) for $\alpha = x, z$ (perpendicular to the cylindrical axis). Note that the NRET rate for the NW→QW case follows the well-known asymptotic behavior $\gamma \propto d^{-4}$ akin to the NP→QW and QW→QW cases reported in [40] and [41], respectively. Figure 5(b) shows the average NRET rate for a CdTe D-A pair as a function of the distance, when the donor is an NW and the acceptor is a QW. In this computation, we made similar assumptions as in the previous section.

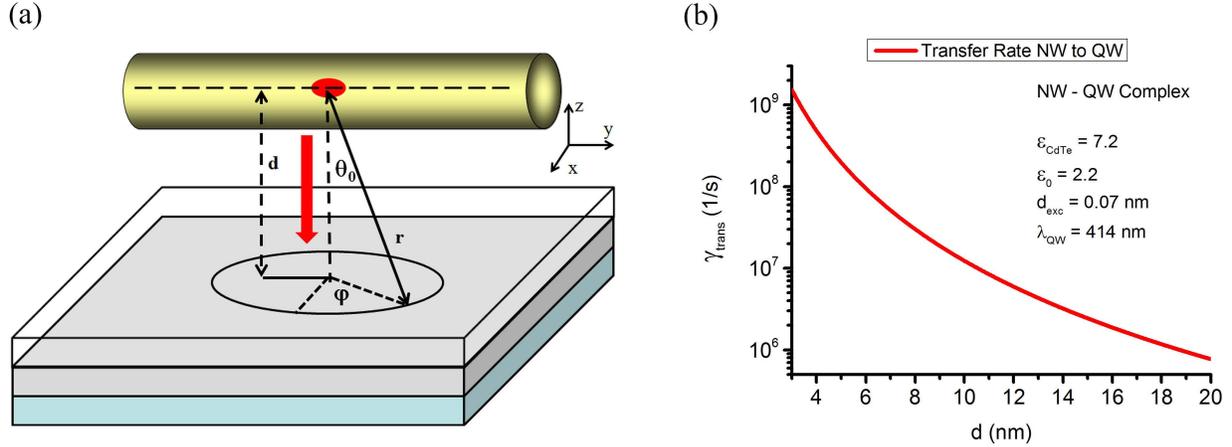

**Figure 5:** a) Schematic for the energy transfer of NW→QW. Red arrows show the energy transfer direction. Red circles represent an exciton in the *α*-direction. *d* is the separation distance. $\theta_0$ is the azimuthal angle between **d** and **r**. $\varphi$ is the radial angle. b) Average NRET rate for a CdTe D-A pair. This plot shows the distance dependency of the NRET rate for the NW→QW case.

To summarize our NRET studies, Table 2 lists the transfer rates in the long distance asymptotic behavior in the dipole approximation for all possible combinations with mixed dimensionality. Here, Table 2 illustrates the functional distance dependency for the NRET: (1) when the acceptor is an NP, NRET is inversely proportional to $d^6$ (Eq. (32)); (2) when the acceptor is an NW, NRET is proportional to $d^{-5}$ (Eq. (40)); and (3) when the acceptor is a QW, NRET is proportional to $d^{-4}$ (Eq. (44)). This indicates that the donor dimensionality (NP, NW, QW) does not affect the functional dependency on the distance. To complete our analysis, we also graphically present the distance dependencies, given in Table 2, in Figure 6 (top panel). Here, the energy transfer rates are given as a function of $d/d_0$, where $d_0$ is the characteristic distance, which satisfies the asymptotic condition required for each case $\left(d \gg R_{NP,(NW)}, d \gg L_{QW}\right)$. Figure 6 (bottom panel) further illustrates the energy transfer efficiency for the NRET as a function of $d/d_0$. As expected, the fastest decay is possible when the acceptor is an NP and the slowest decay is possible when the acceptor is a QW. In all cases, the NRET's distance dependency is given by the acceptor geometry and it is independent of the donor's geometry. Note that the effective dielectric constant, however, depends only on the donor's geometry. Therefore, we can conclude that the NRET's distance dependency is dictated by the confinement degree of the acceptor nanostructure whereas the donor's confinement affects the modification of effective dielectric constant.

**Table 2:** NRET rate summary for the long distance asymptotic limit. This list shows the distance dependence of the NRET rate given the acceptor's geometry. Red color indicates the cases that have not been theoretically studied before.

| Donor \ Acceptor | NP | NW | QW |
|---|---|---|---|
| NP | NP→NP | NP→NW | NP→QW |
| NW | <span style="color:red">NW→NP</span> | NW→NW | <span style="color:red">NW→QW</span> |
| QW | <span style="color:red">QW→NP</span> | <span style="color:red">QW→NW</span> | QW→QW |
| Acceptor Distance Dependency | $\gamma_{NP} \propto \dfrac{1}{d^6}$ | $\gamma_{NW} \propto \dfrac{1}{d^5}$ | $\gamma_{QW} \propto \dfrac{1}{d^4}$ |

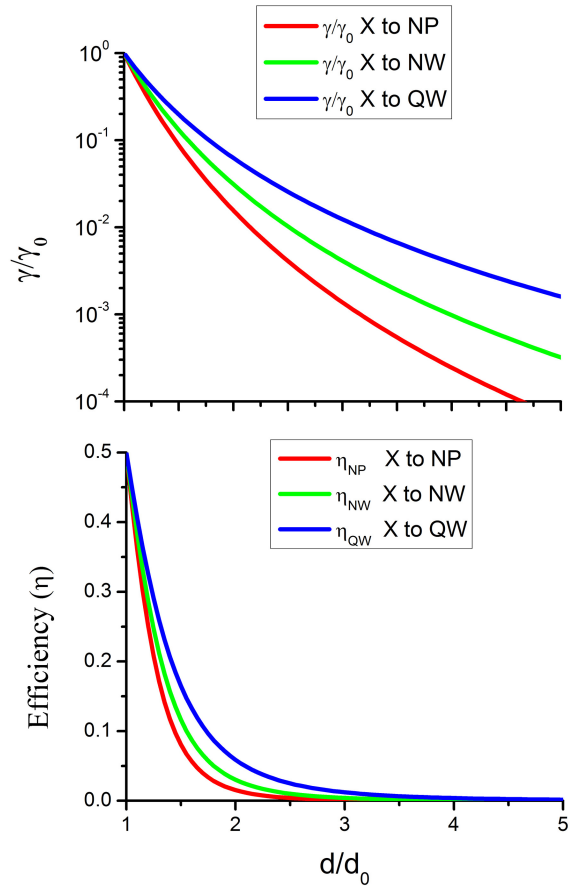

**Figure 6:** NRET rate distance dependency in the long distance asymptotic limit (top). Energy transfer rates and efficiencies are plotted as a function of $d/d_0$, where $d_0$ is the characteristic distance, which satisfies the asymptotic condition required for each case $(d \gg R_{NP,(NW)}, d \gg L_{QW})$. Red line shows the case when the acceptor is an NP. Green line illustrates the case when the acceptor is an NW. Blue line depicts case when the acceptor is a QW. In all of them X can be an NP, an NW, or a QW. Energy transfer efficiency for the NRET in the long distance asymptotic limit (bottom).

# VII. CONCLUSION

In summary, we have obtained a unified picture and understanding of the nonradiative energy transfer for nanostructures of mixed dimensionality, and completed the missing cases (NW→NP, QW→NP, QW→NW, and NW→QW) for the NRET rates in all possible combinations. We obtained analytical expressions for the NRET rate in the long distance approximation and compared our results with the ones already reported in the literature and included those cases that have not previously been studied. We showed that the distance dependence of the NRET rate depends on the geometry and dimensionality of the acceptor and on the effective dielectric constant of the donor. The expressions presented in this work will be a convenience reference to estimate the NRET rate for nanostructures involving mixed dimensionalities. In addition, the NRET results obtained here can help in the optimization and the design of new experiments and/or new devices for high efficiency light harvesting and light generation systems.

# ACKNOWLEDGMENT

This work is supported by National Research Foundation of Singapore under NRF-CRP-6-2010-02 and NRF-RF-2009-09. H.V.D. also acknowledges support from ESF-EURYI and TUBA-GEBIP. A.O.G. acknowledges support from NSF (USA), Volkswagen Foundation (Germany) and Air Force Research Laboratories (Dayton, OH, USA).